\documentclass[a4paper, 10pt, twocolumn]{article}

\usepackage{authblk}
\usepackage{graphicx}
\usepackage{enumerate}
\usepackage[cmex10]{amsmath}
\usepackage{amsthm}
\usepackage{amssymb}

\newtheorem{theorem}{Theorem}
\newtheorem{proposition}{Proposition}

\newtheorem{definition}{Definition}
\usepackage{hyperref}

\title{Strategic Resource Allocation for Competitive Influence in Social Networks}
\author[1]{Antonia Maria Masucci
\thanks{Email: \href{mailto:antonia.masucci@inria.fr}{antonia.masucci@inria.fr}}}
\affil[1]{INRIA Paris-Rocquencourt\\ Domaine de Voluceau B.P. 105\\ 78153 Le Chesnay\\ France}
\author[2]{Alonso Silva
\thanks{Email: \href{mailto:alonso.silva@alcatel-lucent.com}{alonso.silva@alcatel-lucent.com}
To whom correspondence should be addressed.}}
\affil[2]{Alcatel-Lucent Bell Labs France\\ Centre de Villarceaux\\ Route de Villejust\\ 91620 Nozay\\ France}
\date{}

\begin{document}

\maketitle
\begin{abstract}
One of the main objectives
of data mining is to help
companies determine to which
potential customers to market
and how many resources to allocate
to these potential customers.
Most previous works on
competitive influence in social networks
focus on the first issue.
In this work, our focus is on
the second issue, i.e., 
we are interested on  
the competitive influence
of marketing campaigns
who need to simultaneously
decide how many resources to
allocate to their potential customers
to advertise their products.
Using results from game theory,
we are able to completely characterize
the optimal strategic resource allocation
for the voter model of social networks
and prove that the price of competition
of this game is unbounded.
This work is a step towards providing
a solid foundation for marketing advertising
in more general scenarios.
\end{abstract}

\section{Introduction}\label{sec:introduction}

In contrast to mass marketing,
where a product is promoted
indiscriminately to all potential customers,
direct marketing promotes a product only to customers
likely to be profitable.
The groundbreaking works of Domingos and Richardson~\cite{DomingosR2001, RichardsonD2002}
incorporated the influence of peers on the
decision making process of potential customers
deciding between different products or services promoted by 
competing marketing campaigns through direct marketing.
This aggregated value of a customer has been called the network value of a customer.
If each customer was making a buying decision
independently of all other customers,
then we should only consider the intrinsic
value of a customer (i.e.,
the expected profit from sales to him).
However, an individual's decision
to buy a product or service
is often strongly influenced
by his friends, acquaintances, etc.
A customer whose intrinsic value
is lower than the cost of marketing
may be worth marketing to
when his network value is considered.
Conversely, marketing to a profitable
consumer may be redundant if network
effects already make him
very likely to buy~\cite{DomingosR2001}.

Most of the existing literature assumes there is
an incumbent that holds the market 
and a challenger who needs to allocate
advertisement through direct marketing
for certain individuals in order to promote
the challenger product or service.
Notable exceptions to that trend are the works
on competitive influence in social networks
of Bharathi et al.~\cite{BharathiKS2007},
Sanjeev and Kearns~\cite{GoyalK2012},
He and Kempe~\cite{HeK2013},
Borodin et al.~\cite{BorodinFO2010}
and Chasparis and Shamma~\cite{ChasparisS2010}.
Bharathi et al.~\cite{BharathiKS2007} proposed a generalization of the independent
cascade model~\cite{GoldenbergLM2001}
and gave a $(1-1/e)$ approximation
algorithm for computing the best response
to an already known opponent's strategy
and proved that the price of competition
of the game
(resulting from the lack of coordination
among the agents)
is at most $2$.
Sanjeev and Kearns~\cite{GoyalK2012}
considered two independent functions
denoted switching function and
selection function.
The switching function 
takes into account the probability of a consumer
switching from non-adoption
to adoption 
and the selection function
specifies, conditional on switching,
the probability
that the consumer
adopts one of them.
Both players simultaneously choose
some number of nodes to initially seed.
The authors 
make the simplifying assumption
that once a node is infected,
it never switches again, and
proceed to study some specific families of
switching and adoption functions.
He and Kempe~\cite{HeK2013}
motivated by~\cite{GoyalK2012}
studied the price of anarchy of that
framework and found an upper bound
of $2$ on that price.
Borodin et al.~\cite{BorodinFO2010}
showed that for a broad family of competitive
influence models is NP-hard
to achieve an approximation that
is better that the square root
of the optimal solution.
Chasparis and Shamma~\cite{ChasparisS2010}
found optimal advertising policies using dynamic programming
based on the models of~\cite{DubeyGM2006} and~\cite{Friedkin2001}.

In the present work, 
our focus is different from
previously described works where
the focus was to which potential
customers to market,
assuming that we have knowledge
about the cost of adoption of
potential customers,
while the focus in our work
is on how many resources to allocate
to potential customers for them
to prefer one product or service
versus another.
We are interested
on the scenario when two competing
marketing campaigns
need to simultaneously
decide how many resources to allocate
to potential customers to advertise
their products.
The process and dynamics by which
influence is spread is given
by the voter model.

Under that model, our main results are the following:

\begin{theorem}\label{theo:symmetric1}
Given a graph~$G=(V,E)$ representing a social
network of $n$ potential customers.
The symmetric strategic resource allocation problem
by only taking into account the intrinsic value of the customers,
for any target time~$\tau$,
is given by a probability distribution function $F^*$
of $x\in\Delta^{n-1}$, such that each vector
coordinate $x_i$ is uniformly distributed on $[0,2B/n]$
for $i\in\{1,\ldots,n\}$, where $B$ is the available budget
and $\Delta^{n-1}$ is the 
set of available allocations.
\end{theorem}

\begin{theorem}\label{theo:symmetric2}
Given a graph~$G=(V,E)$ representing a social
network of $n$ potential customers.
The symmetric strategic resource allocation problem
for target time~$\tau$ has a solution given
by a probability distribution function $F_\tau^*$
of $x\in\Delta^{n-1}$, such that
each vector coordinate $x_i$
is uniformly distributed on $[0,2B\sum_{j=1}^n M^\tau (i,j)]$
for $i\in\{1,\ldots,n\}$, where $B$ is the available budget,
$M$ is the normalized transition matrix
and $\Delta^{n-1}$ is the 
set of available allocations.
Moreover, the long term case has a solution given
by a probability distribution function $F_\infty^*$
of $x\in\Delta^{n-1}$, such that
each vector coordinate $x_i$
is uniformly distributed on $[0,B d_i/\lvert E\rvert]$
for $i\in\{1,\ldots,n\}$, where $B$ is the available budget,
$d_i$ is the degree of potential customer~$i$,
$\lvert E\rvert$ is the total number of edges of the graph
and $\Delta^{n-1}$ is the 
set of available allocations.
\end{theorem}

\begin{theorem}\label{theo:pricecompetition}
The price of competition of the game
(resulting from lack of coordination among the
agents) is unbounded.
\end{theorem}

Theorem~\ref{theo:symmetric1} gives an
optimal policy for the allocation of resources
in the case when we consider potential customers in isolation
or their intrinsic value,
while Theorem~\ref{theo:symmetric2} gives
an optimal policy for the allocation of resources
in the case when we also include the network
value of potential customers.
Theorem~\ref{theo:pricecompetition}
gives the price of competition
(resulting from the lack of coordination among the agents)
for the strategic allocation problem.
We notice that this is a very different result than
the one obtained by Bharathi et al.~\cite{BharathiKS2007}
where they found that the price of competition is at most
a factor of $2$ when the focus is on which potential customers to recruit.

\subsection{Related Work}
The (meta) problem of influence maximization
was first defined by Domingos and Richardson~\cite{DomingosR2001,RichardsonD2002}.
They studied this problem in a probabilistic setting and
provided heuristics to compute an influence maximizing set.
Following this work, Kempe et al.~\cite{Kempe2003,Kempe2005}
and Mossel and Roch~\cite{MosselR2007}, based on
the results of Nemhauser et al.~\cite{NemhauserWF1978}
and Vetta~\cite{Vetta2002}, proved that for
very natural activation functions (monotone and submodular 
or in economic terms, with decreasing marginal utility),
the function of the expected number of active nodes
at termination is a submodular function
and thus can be approximated through
a greedy approach with a $(1-1/e-\varepsilon)$-approximation
algorithm for 
the spread maximization set problem.
A slightly different model but with similar flavor,
the voter model, was introduced by Clifford and Sudbury~\cite{CliffordS1973}
and Holley and Liggett~\cite{HolleyL1975}.
In that model of social network, Even-Dar and Shapira~\cite{EvenDar2007}
found an exact solution to the spread
maximization set problem when all the nodes
have the same cost and provided an FPTAS
(Fully Polynomial Time Approximation
Scheme is an algorithm
that for any $\varepsilon$ approximates
the optimal solution up to an error $(1+\varepsilon)$
in time $\mathrm{poly}(n/\varepsilon)$)
for the case in which different nodes
may have different costs.

In this work, we study the case when two marketing campaigns
competing to promote a product or service
need to decide how many resources to allocate to potential customers
to advertise
their product or service.
Other works related to competitive influence
where the already 
described~\cite{BharathiKS2007,BorodinFO2010,ChasparisS2010,GoyalK2012,HeK2013}.

To study this case, we use recent advances
of game theory, and in particular of Colonel Blotto games.
In the simplest version of the Colonel Blotto game,
two generals want to capture three equally valued
battlefields. Each general disposes of one
divisible unit of military resources.
The generals have to simultaneously
allocate these resources across the three battlefields.
A battlefield is captured by a general if he allocates
more resources there than his opponent.
The goal of each general is to maximize the
number of captured battlefields.

The relationship between Colonel Blotto games and our work is the following.
We establish a parallel between the
marketing campaigns and the generals;
and between the potential customers and
the battlefields.
Each marketing campaign needs to
strategically allocate advertising resources
to outperform the competing marketing campaign.
This needs to be done while knowing that
the competing marketing campaign is
trying to do the same.
It is thus a typical situation
in which game theory comes into play.
In our case, of course, we will not be
dealing with three potential customers
so we need to extend this case to
include any number of potential customers.
By including the network value of customers,
each potential customer will not be equally
valued so we will need also to consider
different payoffs for different potential customers.

The Colonel Blotto game,
was first solved for the case of three battlefields by Borel~\cite{Borel1921,BorelV1938}.
For the case of equally valued battlefields,
also known as homogeneous battlefields case,
this result was generalized for any number of battlefields by Gross and Wagner~\cite{GrossW1950}.
Roberson~\cite{Roberson2006}
focused on the case of homogeneous battlefields and different budgets (also known as asymmetric budgets case).
Gross~\cite{Gross1950}
proved the existence and a method to construct the joint probability distribution.
Laslier and Picard~\cite{LaslierP2002} and Thomas~\cite{Thomas2013} provided
alternative methods to construct the joint distribution by extending the disc method
proposed by Gross and Wagner~\cite{GrossW1950}.
We will extensively use the work of Gross~\cite{Gross1950}
to derive our results.

The plan of this work is as follows.
In Section~\ref{sec:some-game-theory},
we provide some preliminaries about
the tools we use to find
the optimal strategy.
The reader acquainted with game theory
and Colonel Blotto games can skip
this section.
In Section~\ref{sec:voter-model},
we explain the voter model of social networks.
In Section~\ref{sec:results},
we derive the main results of this work,
and in Section~\ref{sec:conclusions},
we conclude this work and
provide future perspectives
for continuing our work.

\section{Preliminaries on game theory}\label{sec:some-game-theory}

Colonel Blotto games (or Divide a Dollar games)
in their classic version
are a class of two-person zero-sum games,
in which both players need to simultaneously
allocate limited resources over several objects.
Colonel Blotto games are usually described in a military context
where the limited resources are troops and
the objects are battlefields.
In that context, the player devoting
the most troops (or resources) to a battlefield (or object) captures
that battlefield (and the payoff associated with that battlefield)
and its total payoff is the sum of the individual payoffs
across captured battlefields.
To analyze these games, we need to introduce some basic
concepts in game theory.

In this subsection, we follow the notation of~\cite{Vetta2002}.
Consider we have two agents (or players) and disjoint groundsets~$V_1$ and $V_2$.
Each element in $V_i$ represents an act that agent $i$ may make, $i\in\{1,2\}$,
in our case, the allocation of troops to a battlefield.
Let $a_i\subseteq V_i$ be an action (set of acts)
available to agent $i$, for example,
the set of troops allocations across battlefields.
We want to restrict the set of actions
an agent may make; thus we may not allow every subset of $V_i$
to be a feasible action.
We let $\mathcal{A}_i=\{a_i\subseteq V_i: a_i~\textrm{is a feasible action}\}$
be the set of all available actions to agent~$i$.
In our case, the set of possible allocations is limited by the budget of the agents.
We call $\mathcal{A}_i$ the action space for agent $i$.
A {\em pure strategy} is one in which the agent decides to carry out
a specific action.
A {\em mixed strategy} is one in which the agent decides upon
an action according to some probability distribution.
The strategy space $\mathcal{S}_i$ of agent $i$ is the set of mixed strategies.
We let $\mathcal{A}=\mathcal{A}_1\times\mathcal{A}_2$,
$\mathcal{S}=\mathcal{S}_1\times\mathcal{S}_2$,
and $V=V_1\cup V_2$.
Given an action set $A\in\mathcal{A}$,
let $A\oplus a_i'$ denote the action set obtained
if agent $i$ changes its action from $a_i$ to $a_i'$.
Similarly, given a strategy set $S\in\mathcal{S}$,
let $S\oplus s_i'$ denote the strategy set obtained
if agent $i$ changes its strategy from $s_i$ to $s_i'$.
For every agent $i$, there is a private utility function
$\alpha_i:2^{V}\to\mathbb{R}$.
The expected value of $\alpha_i(S)$ on
the strategy set~$S$, denoted $\bar\alpha_i(S)$,
is given by
\begin{equation*}
\bar\alpha_i(S)=\sum_{A\in\mathcal{A}}\alpha_i(A)\mathbb{P}(A|S),
\end{equation*}
where $\mathbb{P}(A|S)$ is the probability that action
set $A$ is implemented given that the agents are using
the strategy set $S$.
The goal of each agent is to maximize its expected private utility.
We say that a set of strategies $S\in\mathcal{S}$ is a {\sl Nash equilibrium}
if no agent has an incentive to change strategy. That is,
for any agent $i$,
\begin{equation*}
\bar\alpha_i(S)\ge\bar\alpha_i(S\oplus s_i')\quad\forall s_i'\in S_i.
\end{equation*}
We say that a Nash equilibrium is a pure strategy Nash equilibrium
if, for each agent $i$, $s_i$ is a pure strategy.
Otherwise, we say that the Nash equilibrium
is a mixed strategy Nash equilibrium.
Throughout this work the term equilibrium
refers to Nash equilibrium, although,
since the game is a zero sum game,
these equilibrium strategies are also optimal strategies.

The following result, due to Nash~\cite{Nash1951},
shows that there exists at least one Nash equilibrium
for any finite game.
\begin{theorem}[Nash~\cite{Nash1951}]
Any finite, $k$-person,
non-co- operative game has
at least one Nash equilibrium.
\end{theorem}

\subsection{Colonel Blotto games}

In this subsection, we include the notation of~\cite{Thomas2013}.
As previously said in Section~\ref{sec:introduction},
in the simplest version of the Colonel Blotto game,
two generals want to capture three equally valued
battlefields. Each general disposes of one
divisible unit of military resources
and the generals have to simultaneously
allocate these resources across the three battlefields.
A battlefield is captured by a general if he allocates
more resources there than his opponent
and the goal of each general is to maximize the
number of captured battlefields.

In that game, a pure strategy for a player, denoted~$X$, is a
$3$-dimensional allocation vector ${\bf x}=(x_1,x_2,x_3)$
where $x_i$ is the amount of resources allocated to
the $i$th battlefield for $i\in\{1,2,3\}$. The set of pure strategies
is the $2$-dimensional unit simplex 
\begin{equation*}
\Delta^2=\{(x_1,x_2,x_3): x_1,x_2,x_3\ge0\quad\textrm{and}\quad x_1+x_2+x_3=1\}.
\end{equation*}
A mixed strategy is a tri-variate distribution
function\break \mbox{$F:\Delta^2\to[0,1]$}.
Similarly for his enemy, denoted~$Y$,
we define ${\bf y}$ the allocation vector, $y_i$ the proportion
of resources allocated to the $i$th battlefield for $i\in\{1,2,3\}$,
$\Delta^2$ the set of pure strategies and \mbox{$G:\Delta^2\to[0,1]$}
a mixed strategy.

The natural extension of the classic version of the Colonel Blotto game
is to study the case with $n$ battlefields
where each captured battlefield gives a different payoff.
Consider that we have two players, denoted~$X$ and $Y$, and $n$ objects.
Player $X$ has budget $B_X$ and it can allocate
for an object~$i$ a proportion of his budget~$x_i$ for \mbox{$1\le i\le n$}.
Similarly, player $Y$ has budget $B_Y$ and it can allocate
for an object~$i$ a proportion of his budget~$y_i$ for $1\le i\le n$.
In this work, we limit ourselves to the case when both players
have the same total divisible budget,
i.e., $B_X=B_Y$, and without loss of generality we consider
this budget to be equal to~$1$.
We consider this to simplify the derivations,
however it is easy to derive the results by considering 
the budgets to be equal to~$B$ instead of~$1$.

For the general case, a pure strategy for player $X$ can be written as an
$n$-dimensional allocation vector\break
${\bf x}=(x_1,x_2,\ldots,x_n)$ with
\begin{equation*}
\sum_{i=1}^n x_i=1, x_i\in [0,1],
\end{equation*}
where $x_i$ represents the fraction of budget allocated
to front~$i$.
Thus, the set of pure strategies
is the $(n-1)$-dimensional simplex
\begin{equation*}
\Delta^{(n-1)}=\{(x_1,\ldots,x_n):x_i\ge0,\,1\le i\le n\  
\textrm{and}\  \sum_{i=1}^nx_i=1\}.
\end{equation*}
A mixed strategy is an $n$-variate distribution
function\break \mbox{$F:\Delta^{(n-1)}\to[0,1]$}.
Let $F_i$ denote the $i$th one-dimensional
marginal of $F$, i.e., the
unconditional distribution of $x_i$.

The object~$i$ has an associated non-negative payoff $A_i$ for $1\le i\le n$.
We denote the sum of the payoffs of all objects by $A$, i.e.,
\begin{equation*}
A=\sum_{i=1}^n A_i.
\end{equation*}
For all $i$, let us define
\begin{equation*}
a_i=\frac{A_i}{A},
\end{equation*}
which represents the relative value of object $i$
and note that
\begin{equation*}
\sum_{k=1}^n a_i = 1.
\end{equation*}

We assume that the player devoting the most resources
to a battlefield captures that battlefield.
Ties are resolved by flipping a coin.

For any pair $({\bf x}, {\bf y})$ of pure strategies,
the excess aggregate value 
for player~$X$, denoted by $g({\bf x},{\bf y})$, of objects captured by player $X$
if he plays the pure strategy ${\bf x}$ while player $Y$ plays the pure strategy ${\bf y}$
is given by
\begin{equation}\label{eq:excess-value}
g({\bf x},{\bf y})=\sum_{i=1}^n a_i\mathrm{sgn}(x_i-y_i),
\end{equation}
where $\mathrm{sgn}(\cdot)$ is the sign function defined as
\begin{equation}\label{eq:sign}
\mathrm{sgn}(u)=
\left\{
\begin{array}{rl}
1 & \textrm{if } u>0,\\
0 & \textrm{if } u=0,\\
-1 & \textrm{if } u<0.
\end{array}
\right.
\end{equation}

The excess aggregate value $g({\bf x},{\bf y})$
is the gain for pure strategy ${\bf x}$
against pure strategy ${\bf y}$.

\begin{definition}
We define the Colonel Blotto game
as the two-player, zero-sum game
defined by the payoff function $g$.
\end{definition}

If $F$ and $G$ are two mixed strategies
the payoff to mixed strategy $F$ against mixed strategy $G$ is:
\begin{equation}\label{eq:mixed}
K(F,G)=\int_{{\bf x}\in\Delta^{(n-1)}}\int_{{\bf y}\in\Delta^{(n-1)}}
g({\bf x},{\bf y})\, dF({\bf x})\, dG({\bf y}).
\end{equation}

The expected payoff for mixed strategy $F$
against pure strategy ${\bf y}$ is given by
\begin{equation}\label{eq:pure}
K({\bf y})=\int_{{\bf x}\in\Delta^{(n-1)}} g({\bf x},{\bf y})\,dF({\bf x}).
\end{equation}
The game is symmetric so to prove that a strategy is optimal
we only need to show that $K({\bf y})\ge0$ for every ${\bf y}$.

From eq.~\eqref{eq:excess-value} and
eq.~\eqref{eq:pure}, we have that
\begin{equation}\label{eq:doumbodo}
K({\bf y})=\sum_{i=1}^n a_i\left(\mathbb{P}(x_i>y_i)-\mathbb{P}(y_i>x_i)\right).
\end{equation}

We assume $n\ge 2$ otherwise the game always ends in a tie.

We observe the following:
\begin{enumerate}[(a)]
\item For the case $n=2$ (originally solved in~\cite{GrossW1950}), it is optimal to put all the budget in
the object of maximum value, i.e.,
in $i^*\in\{1,2\}$ such that $a_{i^*}=\max\{a_1,a_2\}$.
In case $a_1=a_2$, choose any of them.
Indeed, without loss of generality assume
the second object gives higher payoff than the first one,
i.e.~$a_2\ge a_1$,
then the expected payoff against a strategy ${\bf y}=(y,1-y)$ will be
\begin{equation*}
a_1\mathrm{sgn}(-y)+a_2\mathrm{sgn}(y)\ge 0,
\end{equation*}
which always gives a non-negative payoff for all \mbox{$y\in[0,1]$}.
\item\label{item:one} For the case when there exists an object~$i$ with
relative value $a_i\ge1/2$, then
the optimal strategy is to put all the budget in
object~$i$.
Indeed, without loss of generality assume
the last object has relative value $a_n\ge1/2$,
then the expected payoff against a strategy ${\bf y}=(y_1,y_2,\ldots,y_n)$ will be
\begin{equation*}
\sum_{i=1}^{n-1}a_i\mathrm{sgn}(-y_i)+a_n\mathrm{sgn}\left(\sum_{i=1}^{n-1} y_i\right)\ge0,
\end{equation*}
which always gives a non-negative payoff.
\item\label{item:two} For the case when there exists an object~$i$ with
relative value $a_i=0$, then from the payoff function,
the optimal strategy is not to put any budget in object~$i$.
\end{enumerate}

From~\eqref{item:one} and~\eqref{item:two},
in the following we assume that
\begin{equation}\label{eq:relative-value}
0<a_i<1/2,\quad\forall i\in\{1,\ldots,n\},
\end{equation}
or equivalently that
\begin{equation*}
0<A_i<\sum_{j\ne i} A_j,\quad\forall i\in\{1,\ldots,n\}.
\end{equation*}

It can be shown that under the previous assumption for $n>2$ there is no pure strategy Nash equilibrium
in the general case.
Indeed, consider a pure strategy ${\bf x}$, select the object of
minimum value where the pure strategy is not zero.
Then ${\bf x}$ 
will lose with respect to the strategy ${\bf y}$
that allocates no resources to the $i$-th object
and more resources
to all the other objects, i.e.,
\begin{equation*}
y_i=0,\quad y_j=x_j+\frac{x_i}{n-1}\quad\forall j\neq i.
\end{equation*}

Therefore, we need to search for
optimal mixed strategies.
For the case of three battlefields,
Gross and Wagner~\cite{GrossW1950}
proved the existence of a mixed strategy solution given as follows.
\begin{theorem}[Gross and Wagner~\cite{GrossW1950}]
For $n=3$, the Colonel Blotto game with heterogeneous
battlefield values has a mixed strategy equilibrium
in which the marginal distribution over front $i$ is uniform
on $[0,2a_i]$ for $i\in\{1,2,3\}$.
\end{theorem}

\begin{figure}[h!]
\centering
\includegraphics[width=0.4\textwidth]{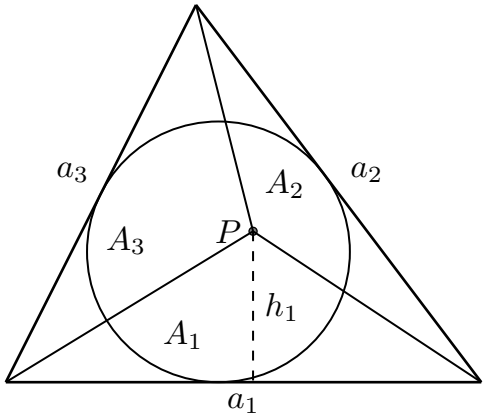}
\caption{Triangle of lengths $a_1$, $a_2$ and $a_3$.}
\label{fig:triangle}
\end{figure}

We need to construct a joint distribution
such that each marginal distribution is uniform
on $[0,2a_i]$ for $i\in\{1,2,3\}$ and such that
the sum of the values given by the marginal distributions is equal to $1$.
The difficulty comes from this last condition,
otherwise we could always define a joint distribution
from its marginal distributions.
In the following, we show how this joint distribution can be constructed.
In the case of three battlefields, there is a geometric construction 
of the joint $n$-variate distribution function with
uniform marginal distribution functions.
\begin{proof}
We construct a non-degenerate triangle 
having sides of length $a_1$, $a_2$ and $a_3$ (see Fig.~\ref{fig:triangle}),
and then inscribe a circle within it
and erect a hemisphere upon this circle.
We notice that we can construct a non-degenerate
triangle since a degenerate triangle would violate 
the assumption given by eq.~\eqref{eq:relative-value}.
Then, we choose a point from a density
uniformly distributed over the surface
of the hemisphere and project this point
straight down into the plane of the triangle
(we denote by~$P$ the projected point over the plane).
We then divide the forces in respective proportion
to the triangular areas subtended by $P$ and the sides,
i.e., $x_1:x_2:x_3=\mathrm{A}_1:\mathrm{A}_2:\mathrm{A}_3$ (see Fig.~\ref{fig:triangle}).

Let $F_i(x_i)$ denote the respective marginal distribution
function of Blotto's continuous mixed strategy $F$.
Blotto's expectation from battlefield $i$ is given by
\begin{equation*}\label{eq:BlottoExpectation}
a_i[1-F_i(y_i)]-a_iF_i(y_i)=a_i[1-2F_i({\bf y})],
\end{equation*}
and hence his total expectation is given by
\begin{equation}\label{eq:total-expectation}
K({\bf y})=\sum_{i=1}^3 a_i[1-2F_i(y_i)].
\end{equation}

Let $h_i$ denote the altitude of the triangle of area $\mathrm{A}_i$
subtended by $P$.
From a well-known property of the surface area
of a sphere, we see that $h_i$ is uniformly
distributed over $(0,2r)$, $r$ being the
radius of the sphere.
Now, $\mathrm{A}_i=\frac12a_ih_i$, and hence
$\mathrm{A}_i$ is uniformly distributed over $(0,a_ir)$.
Also, since $x_1+x_2+x_3=1$ and
$x_1:x_2:x_3=\mathrm{A}_1:\mathrm{A}_2:\mathrm{A}_3$, it follows that
$x_i=\frac{\mathrm{A}_i}{\Delta}$
where $\Delta$ is the area of the originally constructed triangle.
Thus $x_i$ is uniformly distributed over $(0,\frac{a_ir}{\Delta})$.
But $\Delta=\frac12(a_1r+a_2r+a_3r)$,
hence $x_i$ is uniformly distributed over $\left(0,\frac{2a_i}{a_1+a_2+a_3}\right)$,
i.e.,
\begin{equation*}
F_i(x_i)=\min\left[1,\frac{a_1+a_2+a_3}{2a_i}x_i\right].
\end{equation*}

Consequently, from eq.~\eqref{eq:total-expectation},
\begin{equation*}
K(y)=\sum_{i=1}^3 a_i\left\{1-2\min\left[1,\frac{a_1+a_2+a_3}{2a_i}y_i\right]\right\}.
\end{equation*}

Note that if $\alpha>0$, $y\ge0$, then $\min(1,\alpha y)\ge\alpha y$,
and hence
\begin{align*}
K(y)&\ge\sum_{i=1}^3 a_i\left\{1-\frac{a_1+a_2+a_3}{a_i}y_i\right\},\\
&=a_1+a_2+a_3-(a_1+a_2+a_3)\sum_{i=1}^3 y_i\\
&=0
\end{align*}
since $\sum_{i=1}^3 y_i=1$.
\end{proof}

For the case of any number of battlefields~$n$,
Gross~\cite{Gross1950} proved the existence and
a method to construct such joint distributions.

\begin{theorem}[Gross~\cite{Gross1950}]\label{theo:gross}
Consider the Colonel Blotto Game
with heterogeneous battlefield values.
Let $F^*$ be a probability distribution
of ${\bf x}\in\Delta^{n-1}$ such
that each vector coordinate $x_i$
is uniformly distributed on $[0,2a_i]$
for $i\in\{1,\ldots,n\}$.
Then $(F^*,F^*)$ constitutes
a symmetric Nash equilibrium.
\end{theorem}

We notice that Friedman~\cite{Friedman1958}
used this solution for advertisement expenditures
without considering the network value of customers.
Robertson~\cite{Roberson2006}
showed that for homogeneous battlefield
values ($a_i=1/n,\,\forall i\in\{1,\ldots,n\}$)
uniform univariate marginals are also a necessary
condition for equilibrium.
Laslier and Picard~\cite{LaslierP2002}
and Thomas~\cite{Thomas2013}
provided alternative methods to construct the joint
distribution.
In the following we will use Theorem~\ref{theo:gross}
to the strategic resource allocation
on the voter model of social networks.

\section{Voter model}\label{sec:voter-model}

The voter model is one of the most
natural probabilistic models to
represent the diffusion of opinions
in social networks.
In each step, each potential customer
changes its opinion
by choosing one of its
neighbors at random
and adopting its neighbor's opinion.

The voter model is quite different
from the threshold model \cite{Kempe2003,Kempe2005},
however it still has the
same key property that
a person is more likely
to change its opinion
to the one held by most of its neighbors.
It has another characteristic to its advantage,
the fact that in the threshold models
once a node adopts a product or service
it stays with that product or service forever.
These models are called monotone models of diffusion.
However, the voter model allows to change
preferences, which may be more suitable
for non-monotone processes.

Let $G=(V,E)$ be an undirected graph with self-loops
where $V$ is the set of nodes in the graph
which represent the potential customers of
the competing marketing campaigns
and $E$ is the set of edges which
represent the influence between individuals.
We consider that the graph $G$ has $n$ nodes, i.e. $\lvert V\rvert=n$.
As we will see in the following,
the use of the same notation $n$, as for the number of battlefields,
is not a coincidence since each potential customer
will represent a battlefield for competing marketing campaigns.

For a node $v\in V$, we denote by $N(v)$ the set of neighbors
of $v$ in $G$, i.e. $N(v)=\{u\in V: \{u,v\}\in E\}$
and by $d_v$ the degree of node $v$, i.e. $d_v=\lvert N(v)\rvert$.

We recall that the players of the game are the competing
marketing campaigns and the nodes of the graph correspond
to the potential customers.
We label a node $v\in V$ by its initial preference
between different players, $X$ or $Y$,
denoted by function $f_0$.
We denote by $f_0(v)=1$ when node $v\in V$ prefers 
the product promoted by marketing campaign $X$,
$f_0(v)=-1$ when node $v$ prefers
the product promoted by marketing campaign $Y$,
and $f_0(v)=0$ when node $v$ is
indifferent between both products.

We assume that the quantity of marketing budget allocated
to node $i$ determines its initial preference, i.e.,
\begin{equation*}
f_0(i)=
\left\{
\begin{array}{rl}
1 & \textrm{if } x_i>y_i,\\
0 & \textrm{if } x_i=y_i,\\
-1 & \textrm{if } x_i<y_i.
\end{array}
\right.
\end{equation*}
We notice that $f_0(i)$ corresponds to $\mathrm{sgn}(x_i-y_i)$
where $\mathrm{sgn}(\cdot)$ is given by eq.~\eqref{eq:sign}.
We also observe that since the bids are real numbers,
ties have measure zero.
However, in the unlikely event
that bids coincide in the maximum,
we choose uniformly at random the winner
between both players.
This defines a new function and for simplicity we use the same notation,
\begin{equation*}
f_0(i)=
\left\{
\begin{array}{rl}
1 & \textrm{if } x_i>y_i,\\
1 & \textrm{if } x_i=y_i\quad\textrm{and}\quad R=1,\\\
-1 & \textrm{if } x_i=y_i\quad\textrm{and}\quad R=0,\\\
-1 & \textrm{if } x_i<y_i,
\end{array}
\right.
\end{equation*}
where $R$ is a Bernoulli random variable with success probability
equal to $1/2$.

The evolution of the system will be described by the voter model.
Starting from any arbitrary initial preference assignment to the vertices
of $G$, at each time $t\ge 1$, each node picks uniformly at random
one of its neighbors and adopts its opinion.
More formally, starting from any assignment \mbox{$f_0: V\to\{-1,1\}$},
we inductively define
\begin{equation*}
f_{t+1}(v)=
\left\{
\begin{array}{rl}
1 & \textrm{with prob. }\frac{\lvert\{u\in N(v): f_t(u)=1\}\rvert}{\lvert N(v)\rvert},\\
-1 & \textrm{with prob. }\frac{\lvert\{u\in N(v): f_t(u)=-1\}\rvert}{\lvert N(v)\rvert}.\\
\end{array}
\right.
\end{equation*}

The objective function for player $X$ is to maximize
at a certain target time $\tau$ the expected number of nodes:
\begin{equation*}
\mathbb{E}\left[\sum_{v\in V} f_\tau(v)\right].
\end{equation*}

We thus define the strategic resource allocation
problem as follows.

\begin{definition}[Competitive Influence]
Let $G=(V,E)$ be a graph with $n$ nodes
representing a social network with $n$
potential customers.
Consider two players, $X$ and $Y$, which
represent two competing marketing campaigns with equal budget~$B$
which need to simultaneously allocate resources across potential customers.
A node $i\in V$ with respective allocations $x_i$ and $y_i$
such that $x_i>y_i$ (or $y_i>x_i$) will choose the product or
service proposed by player $X$ (or $Y$).
Otherwise, if $x_i=y_i$, it will flip a coin to decide
between both marketing campaigns.
Then the strategic resource allocation problem
is the problem of finding an initial assignment of
resources $x=(x_1,x_2,\ldots,x_n)$
for player $X$ that will maximize
at a target time $\tau$ the
expectation
\begin{equation*}
\mathbb{E}\left[\sum_{v\in V} f_\tau(v)\right]
\end{equation*}
subject to the budget constraint 
\begin{equation*}
\sum_{v\in V}x_v\le B.
\end{equation*}
\end{definition}

\section{Results}\label{sec:results}

In this section,
we establish the main results of our work,
we find the optimal marginal probability
density function when we consider
the intrinsic value of potential customers
and when we consider the total value
of potential customers by incorporating
their network value.
We also give a distance to compare
both probability density functions.

We notice that in the voter model, the probability
that node $v$ adopts the opinion of one its neighbors $u$
is precisely $1/\lvert N(v)\rvert$. Equivalently, this is the probability
that a random walk of length $1$ that starts at $v$
ends up in $u$.
Generalizing this observation by induction on $t$,
we obtain the following proposition.

\begin{proposition}[Even-Dar and Shapira~\cite{EvenDar2007}]
Let $p_{u,v}^t$ denote the probability that a random
walk of length $t$ starting at node $u$
stops at node $v$.
Then the probability that after $t$ iterations of the voter model,
node $u$ will adopt the opinion that node $v$ had at time $t=0$
is precisely $p_{u,v}^t$.
\end{proposition}

Let $M$ be the normalized transition matrix of $G$,
i.e. 
\begin{equation*}
M(v,u)=1/\lvert N(v)\rvert\quad\textrm{if }u\in N(v).
\end{equation*}

By linearity of expectation, we have that for player $X$
\begin{equation*}
\mathbb{E}\left[
\sum_{v\in V} f_\tau(v)\right]=\sum_{v\in V}\left(\mathbb{P}[f_\tau(v)=1]-\mathbb{P}[f_\tau(v)=-1]\right).
\end{equation*}

For a subset $S\subseteq\{1,\ldots,n\}$, we denote by
$1_S$ the 0/1 column vector whose $i$th entry is $1$ if and only
if $i\in S$. Then, the probability that a random walk
of length~$t$ starting at $u$ ends in $v$,
is given by the $(u,v)$ entry of the matrix $M^t$,
or equivalently, by $1_{\{u\}}^TM^t1_{\{v\}}$.
Then
\begin{align*}
\mathbb{P}[f_t(v)=1]&=\sum_{u\in V} p^t_{u,v}\mathbb{P}[f_0(u)=1]\\
&=\sum_{u\in V} 1_{\{u\}}^TM^t1_{\{v\}}\mathbb{P}[x_u>y_u].
\end{align*}

Similarly,
\begin{align*}
\mathbb{P}[f_t(v)=-1]&=
\sum_{u\in V} 1_{\{u\}}^TM^t1_{\{v\}}\mathbb{P}[x_u<y_u].
\end{align*}

Then
\begin{align}\label{eq:bamako}
\mathbb{E}\left[
\sum_{v\in V} f_t(v)
\right]&=\nonumber\\
\sum_{v\in V}\sum_{u\in V} &1_{\{u\}}^TM^t1_{\{v\}}\left(\mathbb{P}[x_u>y_u]-\mathbb{P}[x_u<y_u]\right).
\end{align}

If we are interested only on the
intrinsic value of potential customers,
this case is equivalent to consider that
each node is influenced only by itself, i.e.,
\begin{equation*}
M(u,v)=
\left\{
\begin{array}{cc}
1 & u=v,\\
0 & u\neq v.
\end{array}
\right.
\end{equation*}
Equivalently, that $M=I_{n\times n}$, where $I_{n\times n}$
is the $n\times n$ identity matrix.
Then, for every target time $\tau$ we have that
\begin{equation*}
\mathbb{E}\left[
\sum_{v\in V} f_\tau(v)\right]=\sum_{u\in V} \left(\mathbb{P}[x_u>y_u]-\mathbb{P}[x_u<y_u]\right).
\end{equation*}
From equation~\eqref{eq:doumbodo},
we notice that this case is equivalent to the homogeneous Colonel Blotto,
i.e., with relative values $a_i=1/n$ for $1\le i\le n$,
and from Theorem~\ref{theo:gross}, we are able to conclude Theorem~\ref{theo:symmetric1}.

If we consider a complete graph, then the normalized transition matrix is given by
\begin{equation*}
M=\frac{1}{n}
\left[
\begin{array}{rrrr}
1 & 1 & \ldots & 1\\
1 & 1 & \ldots & 1\\
\vdots & \vdots & \ddots & 1\\
1 & 1 & \ldots & 1\\
\end{array}
\right].
\end{equation*}
We notice that $M^t=M$ for all $t\in\mathbb{N}$.
Thus independently of the target time $\tau$,
the objective function for player $X$ is to maximize:
\begin{equation*}
\mathbb{E}\left[
\sum_{v\in V} f_\tau(v)\right]=\sum_{u\in V} \left(\mathbb{P}[x_u>y_u]-\mathbb{P}[x_u<y_u]\right).
\end{equation*}
Similarly to the previous case, from eq.~\eqref{eq:doumbodo},
we see that this particular case is equivalent to
the homogeneous Colonel Blotto game and the solution
is the same as in the intrinsic value case.

If we could only solve this challenge in the intrinsic case and the complete graph
it would not be a compelling model.
However, as we will see we can compute this when we are interested on much more general cases.

Indeed, from eq.~\eqref{eq:bamako}, redefining
\begin{equation*}
a_u=\sum_{v\in V} M^t(u,v),
\end{equation*}
we have the optimal strategy from Theorem~\ref{theo:gross} and we conclude 
the first part of Theorem~\ref{theo:symmetric2}.

Recall the well known fact that for any graph $G$
with self-loops, a random walk starting from any node~$v$,
converges to the steady state distribution after $O(n^3)$ steps (see~\cite{MotwaniR1995}).
Then the (unique) steady state
distribution is that the probability
of being at node $u$ is $d_u/2\lvert E\rvert$.
In other words, if $t\gg n^3$ then $M_{u,v}^t=(1+o(1))d_u/2\lvert E\rvert$.

If we have $t\gg n^5$, the error in each entry
is within a factor of $1+o(1/n^2)$ of the exact value.
Then
\begin{equation}
K({\bf y})=o\left(\frac1n\right)+\sum_{i=1}^n \frac{d_i}{2\lvert E\rvert}(\mathbb{P}(x_i>y_i)-\mathbb{P}(x_i<y_i)).
\end{equation}

Then, in the long term, i.e., when $t\to+\infty$, from Theorem~\ref{theo:gross}, the optimal strategy
is to play a mixed strategy such that the vector coordinate $x_i$
is uniformly distributed between $0$ and $d_i/\lvert E\rvert$,
which concludes Theorem~\ref{theo:symmetric2}.

\begin{figure}[h!]
\centering
\includegraphics[width=0.4\textwidth]{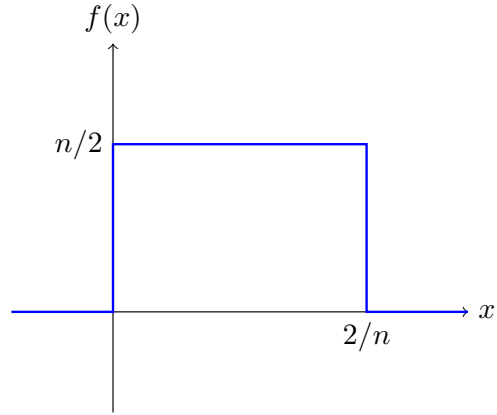}
\caption{Probability density function for the optimal marginal allocation for the intrinsic value case.}
\label{fig:intrinsic-value}
\end{figure}

\begin{figure}[h!]
\centering
\includegraphics[width=0.4\textwidth]{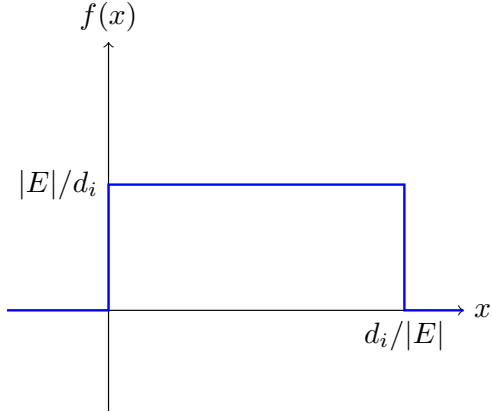}
\caption{Probability density function for the optimal marginal allocation for the total value case.}
\label{fig:total-value}
\end{figure}

\subsection{Distance measure between the intrinsic value and the total value of potential customers}

In this subsection,
we are interested to
find the distance measure between the intrinsic
value of potential customers
and the total value of potential customers. 
From Theorem~\ref{theo:symmetric1},
we know that if we only take into account the intrinsic value
of customers, the optimal strategy for the long term is to choose from
marginal distributions uniform~$x_i^{(1)}\sim\mathcal{U}(0,2/n)$ (see Fig.~\ref{fig:intrinsic-value}).
From Theorem~\ref{theo:symmetric2}, we know that,
by also taking into account the network value of the customers,
the optimal strategy is to choose from
marginal distributions uniform~$x_i^{(2)}\sim\mathcal{U}(0,d_i/\lvert E\rvert)$
(see Fig.~\ref{fig:total-value}).

The marginal distribution of the difference, $Z=x_i^{(1)}-x_i^{(2)}$, is given by:
\begin{itemize}
\item For the case $\frac{d_i}{\lvert E\rvert}\le\frac{2}{n}$:
\begin{equation*}
f_Z(z)=
\left\{
\begin{array}{crl}
\frac{\lvert E\rvert}{d_i}+\frac{n}{2}\frac{\lvert E\rvert}{d_i}z &
-\frac{2}{n}&\!\!\!\!\le z\le\frac{d_i}{\lvert E\rvert}-\frac{2}{n},\\
\frac{n}{2} & \frac{d_i}{\lvert E\rvert}-\frac{2}{n}&\!\!\!\!\le z\le0,\\
-\frac{n}{2}\frac{\lvert E\rvert}{d_i}z+\frac{n}{2} &
0&\!\!\!\!\le z\le\frac{d_i}{\lvert E\rvert}.
\end{array}
\right.
\end{equation*}
\item For the case $\frac{d_i}{\lvert E\rvert}\ge\frac{2}{n}$:
\begin{equation*}
f_Z(z)=
\left\{
\begin{array}{crl}
\frac{\lvert E\rvert}{d_i}+\frac{n}{2}\frac{\lvert E\rvert}{d_i}z &
-\frac{2}{n}&\!\!\!\!\le z\le0,\\
\frac{\lvert E\rvert}{d_i} &
0&\!\!\!\!\le z\le \frac{d_i}{\lvert E\rvert}-\frac{2}{n},\\
-\frac{n}{2}\frac{\lvert E\rvert}{d_i}z+\frac{n}{2} &
\frac{d_i}{\lvert E\rvert}-\frac{2}{n}&\!\!\!\!\le z\le\frac{d_i}{\lvert E\rvert}.
\end{array}
\right.
\end{equation*}
\end{itemize}

We notice that in both cases $f_Z$ is a trapezoidal distribution of mean
\begin{equation*}
\mu_Z=\frac{1}{2}\left(\frac{d_i}{\lvert E\rvert}-\frac{2}{n}\right),
\end{equation*}
and variance
\begin{equation*}
\sigma^2_Z=\frac{1}{12}\left(\left(\frac{2}{n}\right)^2+\left(\frac{d_i}{\lvert E\rvert}\right)^2\right).
\end{equation*}

However, the marginal distribution of the difference is not a
good distance measure since one of the properties that one would
like to have is the identity of indiscernibles, i.e.,
that the distance between two equal probability distributions is zero.
Because of that, we consider the total variation
distance of probability measures. 
\begin{definition}
For $\mu$ and $\nu$ probability measures on $\mathbb{R}$.
The total variation distance between $\mu$ and $\nu$ is defined as
\begin{equation*}
\delta(\mu,\nu)
=\frac12\sup_f\left\lvert\int f(t)\,d\mu(t)-\int f(t)\,d\nu(t)\right\rvert,
\end{equation*}
where the supremum is taken over continuous functions which are bounded by $1$
and vanish at infinity.
\end{definition}

\begin{figure}[h!]
\centering
\includegraphics[width=0.4\textwidth]{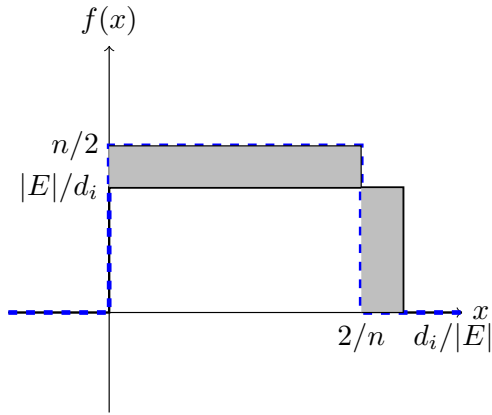}
\caption{Total variation between both probability density functions (shaded region).}
\label{fig:total-variation}
\end{figure}

Informally, the total variation distance is the largest possible
difference between the probabilities that the two
probability distributions can assign to the same event (see Fig.~\ref{fig:total-variation}).

Thus, we have the following two cases:
\begin{itemize}
\item For the case $\frac{d_i}{\lvert E\rvert}\le\frac{2}{n}$:
\begin{equation*}
\delta(x_i^{(1)},x_i^{(2)})=1-\frac{n}{2}\frac{d_i}{\lvert E\rvert}\ge0.
\end{equation*}
\item For the case $\frac{d_i}{\lvert E\rvert}\ge\frac{2}{n}$:
\begin{equation*}
\delta(x_i^{(1)},x_i^{(2)})=1-\frac{\lvert E\rvert}{d_i}\frac{2}{n}\ge0.
\end{equation*}
\end{itemize}

We define the average total variation between both strategies
across all the nodes as
\begin{equation*}
\delta=\frac1n\sum_{i=1}^n\delta(x_i^{(1)},x_i^{(2)}).
\end{equation*}

This distance has some good properties (non-negativity, identity of indiscernibles,
symmetry, triangle inequality).
For the long term,
it can be seen that for any $k$-regular graph, $\delta=0$.
Give the particular probability densities considered,
the effect of $\delta$ will be given by
how much different are $\frac{d_i}{\lvert E\rvert}$
from $\frac{2}{n}$.
For example, for $n\ge2$, a graph with self-loops where all
other edges are given by one node connected
to every other node, there is one node of degree~$(n+1)$
and $(n-1)$ nodes of degree $3$. The total number of edges in the graph
is $2n-1$. Therefore, the average variation is given by
\begin{align*}
&\frac1n\left\{(n-1)\left(1-\frac n2\frac3{2n-1}\right)+\left(1-\frac2n\frac{2n-1}{n+1}\right)\right\}\\
&=\frac{(n-1)(n-2)n(n+1)+2(2n-1)(n-1)(n-2)}{2(2n-1)n^2(n+1)},
\end{align*}
which converges to $1/4$ when $n\to\infty$.
As we have seen, this is very different for other graphs, for example for $k$-regular graphs.

\subsection{Price of competition}

In this subsection, we study the
price of competition in the strategic
resource allocation problem for competitive influence.
We define the price of competition
as the ratio of the total cost to
a competitive duopoly to that of a monopoly.
Informally, it is the price
resulting from the lack of coordination
among the agents~\cite{BharathiKS2007}.

Until now, we have considered
that for any $\epsilon>0$
such that $x_i-y_i\ge\epsilon$
user~$i$ will choose the product
or service of player~$X$ versus
the product or service of player~$Y$.
Under that scenario the price
of competition will be unbounded.
Indeed, the cost for two marketing campaigns cooperating (as in a monopoly)
could put together for each user $\varepsilon/n$
(thus $\varepsilon$ for the whole population),
while for the non-cooperating scenario
each marketing campaign needs to invest a budget
of $B>0$ for the whole population,
thus the price of competition
is $B/\varepsilon$ which is unbounded
for $\varepsilon\to0$.

We notice that this is a very different result than 
the one obtained by Bharathi et al.~\cite{BharathiKS2007}
where they found that the price of competition is at most 
a factor of $2$ when the focus is on which potential customers to recruit.
We have thus obtained Theorem~\ref{theo:pricecompetition}.

\section{Conclusions and Future Work}\label{sec:conclusions}

\subsection{Conclusions}

In this work, we were interested on
the strategic resource allocation
of competing marketing campaigns
who need to simultaneously
decide how many resources to allocate to potential customers
to advertise their products.
Using game theory, and
in particular Colonel Blotto games,
we were able to completely characterize
the optimal strategic allocation of resources
for the voter model of social networks.
We were able to prove that for this case
the price of competition is unbounded.

\subsection{Future Work}

One generalization to several marketing
campaigns consists on the simple case of analyzing
pairwise competitions as previously described.
This case has the advantage that we already know
the solution, given by the results of the previous
section.
However, this is not a realistic case for competitions
in which each customer chooses only one product
from the competing marketing campaigns.
To see this, consider the example of three
competing marketing campaigns $X$, $Y$, and $Z$
and four customers (for the sake of simplification).
Consider the pure strategies
${\bf x}=(0.2,0.2,0.2,0.2,0.2)$,
${\bf y}=(0,0,0,0.5,0.5)$
and ${\bf z}=(0.5,0.5,0,0,0)$.
In that case, the pairwise competition gives
that $X$ has captured $3$ out of $5$ potential customers to $Y$,
and that $X$ has captured $3$ out of $5$ potential customers to $Z$,
thus winning in a pairwise competition against both competitors.
However, since each customer will only choose
one product, the final outcome will be
$2$ customers for $Y$, $2$ customers for $Z$,
while only $1$ customer for $X$.

For exploring this generalization, we need to
search for another framework.
We notice that 
there is a tight relationship between Colonel Blotto games and auctions.
A Colonel Blotto game can be seen as a simultaneous all-pay auction
of multiple items of complete information.
An all-pay auction is an auction in which every bidder
must forfeit its bid regardless of whether it wins the object
which is awarded to the highest bidder.
It is an auction of complete information
since the value of the object is known to every bidder.
In another context, this was already noted by Szentes and Rosenthal~\cite{SzentesR2003},
Roberson~\cite{Roberson2006} and Kvasov~\cite{Kvasov2007}. 

Through this new perspective, we can generalize Colonel Blotto games
to more than two players, where the winner of an object
will be the highest bidder for that object.
We notice that this is a different
generalization than the work of Goldman and Page~\cite{GoldmanP2009}
who consider another payment function.
This general framework does not have yet a known solution
and we consider it an interesting extension to study.

\section*{Acknowledgments}

The work presented in this paper has been partially carried out at LINCS (\url{http://www.lincs.fr}).

\bibliographystyle{hacm}
\bibliography{mybibfile}
\end{document}